\documentclass[showpacs,preprintnumbers,amsmath,amssymb,nofootinbib]{revtex4}
\usepackage{amssymb}
\usepackage{amsfonts}

 \usepackage{epsf}
 \usepackage{graphicx}
\usepackage{float}
\begin{document}
\newcommand{\be}[1]{\begin{equation}\label{#1}}
 \newcommand{\ee}{\end{equation}}
 \newcommand{\bea}{\begin{eqnarray}}
 \newcommand{\eea}{\end{eqnarray}}
 \newcommand{\bed}{\begin{displaymath}}
 \newcommand{\eed}{\end{displaymath}}
 \def\disp{\displaystyle}

 \def\gsim{ \lower .75ex \hbox{$\sim$} \llap{\raise .27ex \hbox{$>$}} }
 \def\lsim{ \lower .75ex \hbox{$\sim$} \llap{\raise .27ex \hbox{$<$}} }

\title{Constructing a cosmological model-independent Hubble diagram of type Ia supernovae with cosmic chronometers}

\author{Zhengxiang Li$^1 \footnote{zxli918@bnu.edu.cn}$, J. E. Gonzalez
        $^2 \footnote{javierernesto@on.br}$, Hongwei Yu$^3 \footnote{hwyu@hunnu.edu.cn}$,
        Zong-Hong Zhu$^1 \footnote{zhzhu@bnu.edu.cn}$, and J. S. Alcaniz$^2 \footnote{alcaniz@on.br}$}

\address{$^1$Department of Astronomy, Beijing Normal University, Beijing 100875, China
\\$^2$Observat\'{o}rio Nacional, 20921-400, Rio de Janeiro, RJ, Brazil
\\$^3$Center of Nonlinear Science and Department of Physics, Ningbo
University,  Ningbo, Zhejiang 315211, China}

\begin{abstract}
We apply two methods, i.e., the Gaussian processes and the non-parametric smoothing
procedure, to reconstruct the Hubble parameter $H(z)$ as a
function of redshift from 15 measurements of the expansion rate
obtained from age estimates of passively evolving galaxies. These
reconstructions enable us to derive the luminosity distance to a
certain redshift $z$, calibrate the light-curve fitting parameters
accounting for the (unknown) intrinsic magnitude of type Ia
supernova (SNe Ia) and construct cosmological model-independent
Hubble diagrams of SNe Ia. In order to test the compatibility
between the reconstructed functions of $H(z)$, we perform a
statistical analysis considering the latest SNe Ia sample, the
so-called JLA compilation. We find that, for the Gaussian processes,
the reconstructed functions of Hubble parameter versus redshift, and thus the following
analysis on SNe Ia calibrations and cosmological implications, are sensitive
to prior mean functions. However, for the non-parametric smoothing method,
the reconstructed functions are not dependent on initial guess models, and consistently
require high values of $H_0$, which are in excellent agreement with recent measurements
of this quantity from Cepheids and other local distance indicators.

\end{abstract}


\maketitle
\renewcommand{\baselinestretch}{1.5}

\section{Introduction}\label{sec1}

The inherent relation between the peak luminosity of type Ia
supernovae (SNe Ia) and the speed of luminosity evolution after
maximum light (known as the Phillips
relationship)~\cite{Phillips1993}, makes possible to standardize
these events as a distance indicator and measure the geometry and
dynamics of the universe. As well known, several years after the
discovery of the Phillips relationship, observations of some dozens
of distant SNe Ia led to the discovery of the cosmic
acceleration~\cite{Riess98,Schmidt98,Perlmutter99}. In Einstein's
general relativity, such behavior implies either the existence of a
new field, the so-called dark energy (see Refs.~\cite{Caldwell10,
Li11} for recent reviews), or that the matter content of the
universe is subject to dissipative processes~\cite{lima_alcaniz,
alcaniz_lima, chimento}. The mysterious cause of the current cosmic
acceleration can also be attributed to a modification of the
standard theory of gravity itself on cosmologically relevant
physical scales~\cite{Ishak06,Kunz07,Bertschinger08}.

In the past decade, several groups have put a lot of effort into
enlarging the sample size of well-measured SNe Ia events. At the
same time, improvements in precision and assessments of systematic
errors have also been accomplished. Recently, several SNe Ia samples
containing a large number of SNe Ia events with high quality have
been released, such as the Sloan Digital Sky Survey-II Supernova
Survey (SDSS-II SN Survey)~\cite{Frieman08,Kessler09}, the Union2
and Union2.1 SN Ia released by the Supernova Cosmological Project
(SCP)~\cite{Amanullah10,Suzuki12}, the first three years of
Supernova Legacy Survey (SNLS3)~\cite{Guy10, Conley11}, and the
joint light-curve analysis (JLA) of SDSS-II and
SNLS3~\cite{Betoule13,Betoule14}.

The distance estimation from SNe Ia data is based on the empirical
observation that these events form a homogeneous class whose
remaining variability is reasonably well captured by two parameters.
One of them describes the time stretching of the light-curve ($x_1$)
whereas the other  describes the SNe Ia color at maximum brightness
($c$). In the latest JLA SNe Ia sample~\cite{Betoule14}, which uses
the SALT2 model to reconstruct light-curve parameters ($x_1$, $c$,
and the observed peak magnitude in rest frame $B$ band
$m_\mathrm{B}^*$), the distance estimator assumes that SNe Ia with
identical color, shape, and galactic environment have on average the
same intrinsic luminosity at all redshifts. This assumption can be
quantified by a linear expression, yielding a standardized distance
modulus, which relates to the luminosity distance $d_\mathrm{L}$ via
$\mu=5\log\big[\frac{d_\mathrm{L}}{\mathrm{Mpc}}\big]+25$:
\begin{equation}\label{eq1}
\mu^{\mathrm{SN}}(\alpha, \beta, M_\mathrm{B})=m_\mathrm{B}^*-M_\mathrm{B}+
\alpha\times x_1-\beta\times c,
\end{equation}
where $\alpha$ and $\beta$ are nuisance parameters which
characterize the stretch-luminosity and color-luminosity
relationships, reflecting the well-known broader-brighter and
bluer-brighter relationships, respectively. The value of
$M_\mathrm{B}$ is another nuisance parameter which represents the
absolute magnitude of a fiducial SNe and was found to depend on the
properties of host galaxies, e.g., the host stellar mass
($M_{\mathrm{stellar}}$), although the reason is not completely
clear~\cite{Sullivan11,Johansson13}. Here, we follow the procedure
in Ref~\cite{Conley11} to approximately correct for this effect by a
simple step function:
\begin{eqnarray}
M_\mathrm{B}=\begin{cases} M_\mathrm{B}^1,~~~&\mathrm{if}~
M_{\mathrm{stellar}}<10^{10} M_\odot.\\
M_\mathrm{B}^1+\Delta_\mathrm{M},~~&\mathrm{otherwise}.
\end{cases}
\end{eqnarray}
In general, the light-curve fitting parameters, $\alpha$ and
$\beta$, are left as free parameters being determined in the global
fit to the Hubble diagram. This treatment results in the dependence
of distance estimation on the cosmological model used in the
analysis. Thus, implications derived from SNe Ia observations with
the light-curve fitting parameters determined in the global fit to
the Hubble diagram are somewhat cosmological-model-dependent.

Our goal in this paper is to construct a completely cosmological
model-independent Hubble diagram of SNe Ia using observational data
of the so-called cosmic chronometers~\cite{Jimenez02}, where the
cosmic expansion rates $H(z)$ are measured from age estimates of red
galaxies without any assumption of cosmology.

\section{Methodology}\label{sec2}

The expansion rate, $H=\dot{a}/{a}$ where $a=1/(1+z)$, at
redshifts $z\neq0$ can be obtained from the derivative of redshift
with respect to cosmic time, i.e.,
$H(z)\simeq-\frac{1}{1+z}\frac{\Delta z}{\Delta t}$. The difficult
task here is to estimate the change in the age of the Universe as a
function of redshift $\Delta t$. This can be done by calculating the
age difference between two luminous red galaxies at different
redshifts, as proposed in Ref.~\cite{Jimenez02}. This method is
usually referred to as differential age and the passively evolving
galaxies from which $\Delta t$ is estimated are called cosmic
chronometers. Currently, 21 measurements of $H(z)$  based on this
method  (in the redshift range $0.070 \leq z \leq 1.965$) have been
obtained~\cite{Jimenez03,Simon05,Stern10,Moresco12,Moresco:2015cya}.
Although cosmological model-independent, these estimates rely on
stellar population synthesis models whose influence on $\Delta t$,
according to Ref.~\cite{morescoetal}, becomes important at $z
\gtrsim 1.2$. In our analysis, we follow Ref.~\cite{licia2014} and
consider only 15 $H(z)$ measurements up to $z < 1.2$ which, in
practice, given the redshift distribution of the $H(z)$ data, means
$z \leq 1.037$. We also increase slightly (20\%) the error bar of
the highest-$z$ point to account for the uncertainties of the
stellar population synthesis models. This ensures that the evolution
of the Hubble parameter as a function of redshift reconstructed in
this paper is neither dependent on the cosmology nor on the stellar
population model.

\subsection{Distances from H(z) measurements}
Recently, $H(z)$ measurements were used to estimate distances by
solving numerically the comoving distance integral for non-uniformly
spaced data with a simple trapezoidal rule~\cite{Holanda13},
\begin{equation}\label{integral}
d_c=c\int_0^z\frac{dz'}{H(z')}\approx\frac{c}{2}\sum_{i=1}^N(z_{i+1}-z_i)
\bigg(\frac{1}{H_{i+1}}+\frac{1}{H_i}\bigg).
\end{equation}
The uncertainty associated to the $i^{th}$ bin is standardly propagated from
the errors of $H(z)$ data,
\begin{equation}
s_i=\frac{c}{2}(z_{i+1}-z_{i})\bigg(\frac{\sigma^2_{H_{i+1}}}{H^4_{i+1}}
+\frac{\sigma^2_{H_{i}}}{H^4_{i}}\bigg),
\end{equation}
and then the error of the integral (\ref{integral}) from $z=0$ to
$z=z_n$ is $\sigma_n^2=\sum^n_{i=1}=s_i^2$.
Naturally, the precision of this simple trapezoidal rule is
sensitive to the uniformity of the spaced data as well as to the
number of data points in a certain spaced range. As indicated in
Ref.~\cite{Kai13}, the relative errors of this method decrease
remarkably when the number of intervals averagely spaced in $z=0-1$
increases. Here, we use two methods to reconstruct the evolution of
the expansion rate with redshift from cosmic chronometer $H(z)$
measurements, namely, Gaussian Processes (GP) and a
non-parametric smoothing (NPS). This procedure
enable us to achieve model-independent distance estimates by
integrating the inverse of the reconstructed function using the
approach of Ref.~\cite{Holanda13} with a very small and uniform step
of $\Delta z=z_{i+1}-z_i$ (see Sec. 3).

\subsubsection{Gaussian processes}\label{sec2.1.1}

As a powerful non-linear interpolating tool, the Gaussian processes
allow us to reconstruct a function from data without assuming a
model or parameterization for it. This method has been first pioneered for
cosmology~\cite{Holsclaw10a,Holsclaw10b}, and then has been widely
used for several purposes, for instance, reconstructions of the equation of state
of dark energy~\cite{Seikel12a}, cosmography~\cite{Shafieloo12d,Shafieloo13d},
null tests of the concordance model~\cite{Seikel12b,Yahya13}, $H_0$ from
cosmic chronometer data~\cite{Busti14}, and reconstructions of the distance-duality
relation~\cite{Yzhang14}.

The reconstruction is based on a mean function with Gaussian error
bands, where the function value at $z$ is not independent of the
function value at some other point $\tilde{z}$ (especially when $z$
and $\tilde{z}$ are close to each other) and they are related
through a covariance function $k(z,\tilde{z})$. This covariance
function depends on a set of hyperparameters and there is a wide
range of possible candidates for it. As the function of Hubble
parameter versus redshift is expected to be infinitely
differentiable, we consider the squared exponential covariance
function:
\begin{equation}\label{eq2}
k(z,\tilde{z})=\sigma_f^2\exp\big\{-\frac{(z-\tilde{z})^2}{2l^2}\big\},
\end{equation}
where the two hyperparameters $\sigma_f$ and $l$ are, respectively,
related to typical changes in the function value and to the length
scale one needs to move in input space to get significant change in
the function value. In order to obtain the value of the function,
the hyperparameters should be trained by maximizing the marginal
likelihood which only depends on the locations of the observations.
As the key steps summarized in Ref.~\cite{Seikel12a}, another issue
in constructing the GP is to decide a prior mean function. In order
to achieve an unbiased reconstruction, we should choose a flat prior
by taking a constant mean function into consideration. Moreover, it
is reasonable and safe to decide a constant function when we do not have
any prior information about the reconstructed result. As it was initiated and
widely used in the literature, the best choice is a constant mean
function because any prior model may introduce significant bias in the results. However,
it is also necessary to check the dependence of the reconstruction and following implications
on the selected prior mean function. Here, we do it by using different best-fit
models, i.e., Einstein-de Sitter (E-D) model, $\Lambda$CDM, and wCDM as the prior mean function.
Results are summarized in Table~\ref{Tab1}.

In this work, we reconstruct the Hubble parameter as a function of
the redshift from 15 $H(z)$ measurements of cosmic chronometers by
using the GaPP (Gaussian Processes in
Python)\footnote{http://www.acgc.uct.ac.za/$\sim$seikel/GAPP/index.html}~\cite{Seikel12a}.

\subsubsection{Non-parametric smoothing method}\label{sec2.1.2}

In order to verify the influence of the reconstructing methods on
the results, we also use the non-parametric procedure of
Ref.~\cite{Shafieloo06,Shafieloo07,Shafieloo10,Shafieloo12} to reconstruct the $H(z)$
function from cosmic chronometer data. The smoothing function
taking into account the data errors is given by
\begin{equation}
H^s(z,\Delta) = H^g(z) + N(z) \sum_i \frac{\left[H(z_i)- H^g(z_i)
\right]} {\sigma^2_{H(z_i)}} \times {\cal{K}}(z, z_i),
\end{equation}
where $H^s(z,\Delta)$, $H^g(z_i)$, $H(z_i)$, $\sigma_H(z_i)$ and
$\Delta$ correspond, respectively, to the smoothed data,  the
initial guess model, the observed data, the error associated with
the $H(z_i)$ data and the smoothing scale. The function $N(z)$ is
the normalization factor given by:
\begin{equation}
 N(z)^{-1}= \sum_i {{\cal{K}}(z, z_i)\over {\sigma^2_{H(z_i)}}}.
\end{equation}
Given the arbitrariness in the choice of the kernel ${\cal{K}}(z,
z_i)$, we tested a Gaussian function and the lognormal kernel of
Ref.~\cite{Shafieloo06}. We found a small difference between the
reconstructed functions from both approaches. In what follows, we
will adopt a Gaussian kernel.

We perform a boot-strapping method applying iterative smoothing
functions. In the first interaction, we subtract an initial guess
model  to the data in order to smooth the noise. Then we add back
the initial guess model. In the next smoothing, we replace
$H^g(z_i)$ by the previous smoothed $H^s(z)$ and calculate the
$\chi^2$ value in each iteration. We stop the  process when the
variation of the $\chi^2$ value between two successive steps is
$\sim 0.002\%$ ($ \simeq 1000^{th}$ iteration).

The reconstruction depends on the value of $\Delta$. For example,
for values of $\Delta < 0.6$ the reconstructed function has many
bumps, for $\Delta$ between $[0.6,0.9]$ bumps disappear but
the reconstruction at high redshift is strongly dependent on the
$\Delta$ value whereas for $\Delta > 1.0$ the reconstruction
in the whole redshift interval considered is weakly dependent
on the $\Delta$ value. However, very high values are meaningless because data
points at high redshift should not be highly correlated with low
redshift data points. In order to select the smoothing scale we
minimize over $\Delta$ the cross-validation function, which is
defined by:

\begin{equation}
 CV(\Delta)=\frac{1}{n}\sum_i (H(z_i)-H^s_{-i}(z_i|\Delta))^2,
\label{CV}
\end{equation}
where $H^s_{-i}(x_i|\Delta)$ denotes the reconstructed Hubble
parameter at $z=z_i$ without taking into account the ponte
$(z_i,H(z_i))$. For our 15 $H(z)$ data discussed in the previous
section the $\Delta$ value that asymptotically minimizes Eq.
(\ref{CV}) is 1.4.

As reported in Ref.~\cite{Shafieloo06}, the initial guess model is
not relevant since the reconstruction does not depend on it. Here, we also have
tested such dependence by considering different best-fit model,
namely, the E-D, $\Lambda$CDM, and wCDM as the initial guess one. Results are summarized in
Table~\ref{Tab2}. An important difference between the
analysis reported here and the one performed in
Ref.~\cite{Shafieloo06} is that in the latter the iteration number
of the boot-strapping process $n$ is the free parameter relative to
which a $\chi^2$ function is minimized to define the $1\sigma$
region ($\chi^2=\chi^2_{min}+1$). In our analysis, in order to
calculate the $1\sigma$ confidence level we extrapolate the
technique developed in~\cite{Bowman_smoothing} for linear smoothing
and consider the expression:
\begin{equation}
 \sigma_{H^s(z)}= \left( \sum_i v^2_i \hat{\sigma}^2 \right)^{1/2},
 \label{1sigma}
\end{equation}
where $\sigma_{H^s(z)}$ is the $1\sigma$ error of the reconstruction,
$v_i$ is the smoothing factor ($v_i=N(z) {\cal{K}}(z,z_i)/\sigma^2_{H(z_i)}$)
and $\hat{\sigma}^2$ is the estimate of the error variance given by
\begin{equation}
 \hat{\sigma}^2=\frac{\sum_i \left(H(z_i)-H^s(z_i) \right)^2}{\sum_j (1-v_j(z_j))}.
\end{equation}
In order to test the validity of this extrapolation, we simulate
$H(z)$ data using the $\Lambda$CDM scenario as fiducial model with
different values of $\sigma$ ($\sigma_{sim}$) and compare them with
those ($\sigma_{rec}$) calculated from Eq. (\ref{1sigma}). We find
that $\sigma_{rec}$ is $\simeq 30\%$ smaller than $\sigma_{rec}$.
We, therefore, add 30\% to the $\sigma_{H^s(z)}$ obtained from Eq.
(\ref{1sigma}).

Figure 1a shows both GP and NPS reconstructions of the expansion
history from the cosmic chronometer data using the best-fit values
for the hyperparameters (GP) and $\Delta$ (NPS). Here, results correspond
to reconstruction with the prior mean function being a constant for GP and
that with the initial guess model being the best-fit $\Lambda$CDM for NPS, respectively.
For completeness, we also plot $H(z)/(1+z)$ as a function of $z$ in Figure 1b. Clearly,
the NPS reconstruction shows a minimum (deceleration/acceleration
transition) at $z_t\sim0.6$ whereas the GP function presents no
transition in the $z$ interval considered. This could be
understood as the latter reconstruction might provide a more reasonable
reconstruction and better calibration to the JLA sample if we believe that the
transition from deceleration to acceleration has happened in the real
case (also in the standard $\Lambda$CDM model).

\begin{figure*}[t]
\label{Fig1}
\includegraphics[width=0.49\textwidth, height=0.45\textwidth]{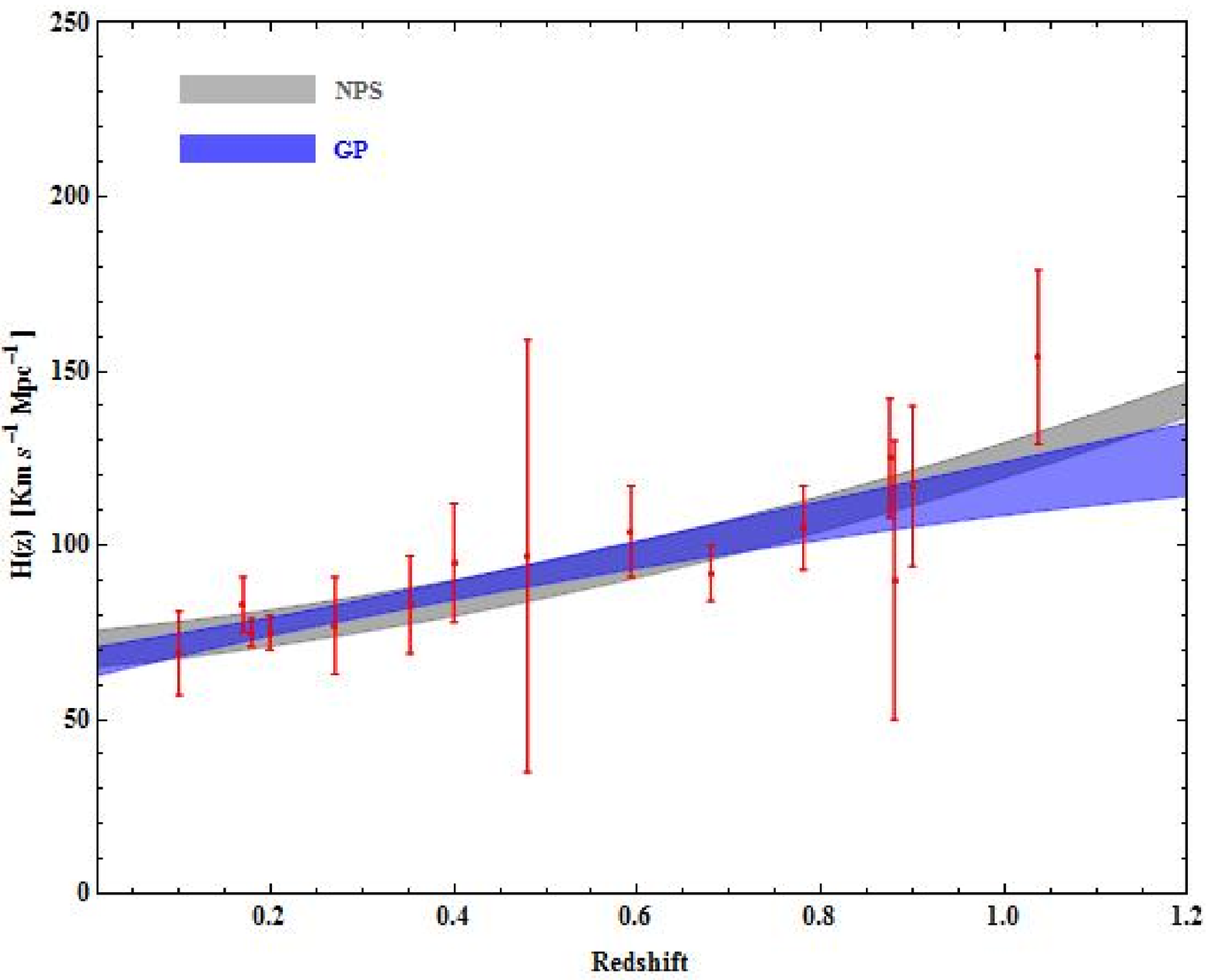}
\includegraphics[width=0.49\textwidth, height=0.45\textwidth]{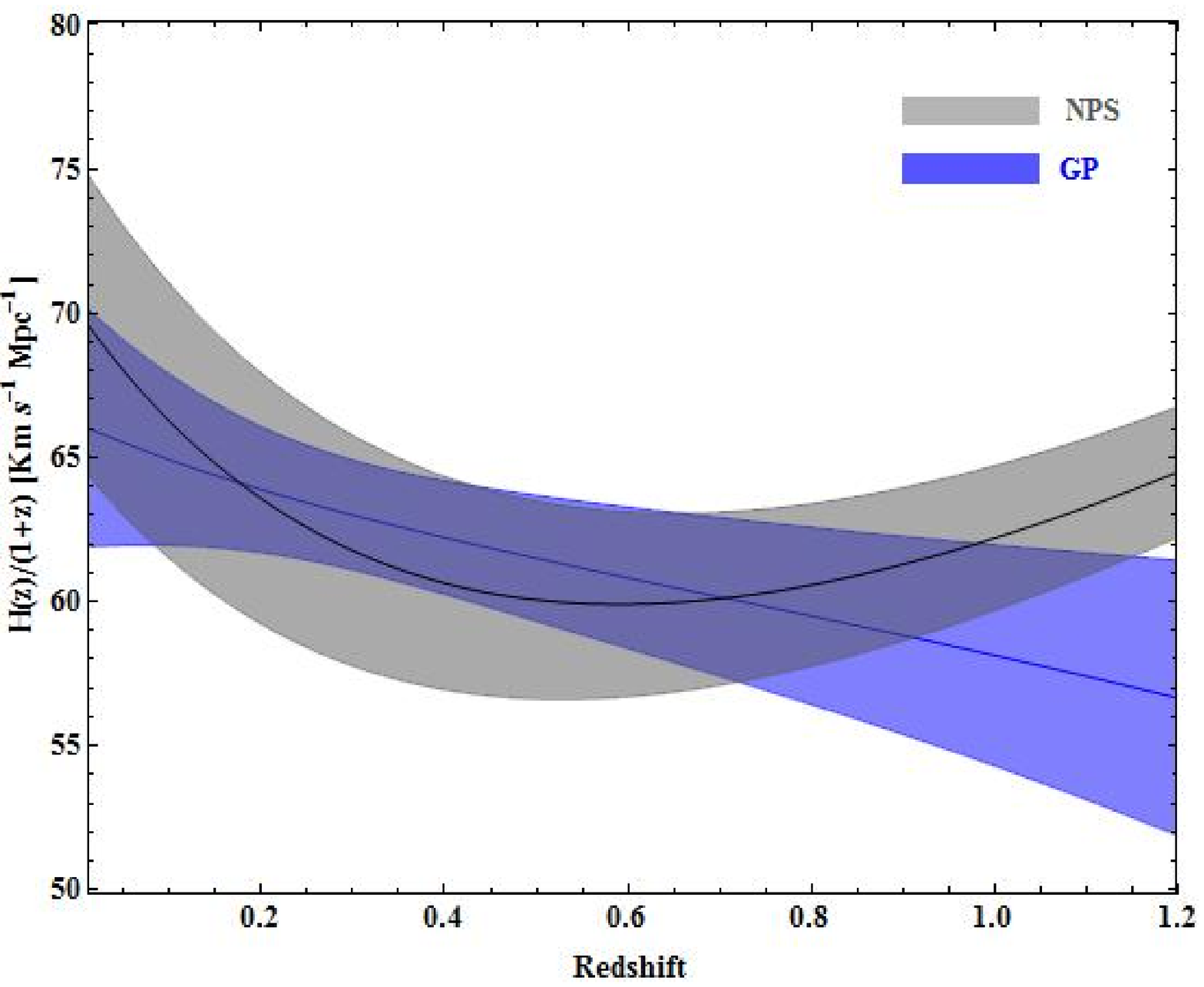}
\caption{ {\emph{a)}} Reconstructions of the
Hubble expansion from cosmic chronometer data using Gaussian
processes (blue region) with the prior mean function being a constant and non-parametric
smoothing method (gray region) with the initial guess model being the best-fit $\Lambda$CDM.
{\emph{b)}} The quantity $H(z)/(1+z)$ as a function of $z$ for both reconstructions.}
\end{figure*}

\subsection{Calibrations for the light-curve fitting parameters}

In order to calibrate the light-curve fitting parameters and
construct cosmological model-independent Hubble diagrams for the JLA
SN Ia sample, we first transform the $H(z)$ reconstructions above into
distance following the approach proposed in Ref.~\cite{Holanda13}.
Assuming a spatially flat universe, we solve numerically the
distance integral and calculate the corresponding uncertainty with a
uniform step, $\Delta z=z_{i+1}-z_i=0.005$. For completeness, we
check the difference between the results from this numerical
treatment and the analytical calculation considering a given
cosmology. At the redshift range considered, $z<1.2$, and assuming
the standard $\Lambda$CDM model, the difference in distance modulus
is $< 0.003$ mag, which is negligible when compared to the
uncertainties of current SNe Ia observations ($\simeq 10^{-1}$ mag).

Since the expansion rate measurements obtained from the derivative of redshift
with respect to cosmic time, i.e., cosmic chronometer, and the non-parametric methods
are cosmology free, distances derived from the reconstructed functions
of Hubble parameter with respect to redshift are considered to be the true ones,
$d_\mathrm{L}^{\mathrm{true}}$ or $\mu^{\mathrm{true}}$ (luminosity distance can be
obtained from the comoving distance by multiplying $(1+z)$: $d_L=(1+z)d_c$). And then,
$\alpha$ and $\beta$ are fitted using the standard minimum-$\chi^2$ route to
analytically marginalize the absolute magnitude of a fiducial SNe Ia, $M_\mathrm{B}$,
\begin{equation}\label{chi2}
\chi^2(\alpha,\beta,M_\mathrm{B})=A-2\times M_\mathrm{B}\times B+M^2_\mathrm{B}\times C,
\end{equation}
where
\begin{eqnarray}\label{ABC}
A(\alpha,\beta)&=&\sum_{i=1}^{740}\frac{[\mu^{\mathrm{SN}}(z_i, \alpha,\beta; M=0)-
\mu_{\mathrm{true}}(z_i)]^2}{\sigma_{\mathrm{tot},i}^2(\alpha, \beta)}\;,\\
B(\alpha,\beta)&=&\sum_{i=1}^{740}\frac{[\mu^{\mathrm{SN}}(z_i, \alpha,\beta; M=0)-
\mu_{\mathrm{true}}(z_i)]}{\sigma_{\mathrm{tot},i}^2(\alpha, \beta)}\;,\\
B(\alpha,\beta)&=&\sum_{i=1}^{740}\frac{1}{\sigma_{\mathrm{tot},i}^2(\alpha,
\beta)}\;.
\end{eqnarray}
Here $\sigma^2_{\mathrm{tot}}$ are propagated from both the statistical uncertainties in
SNe Ia and those in the derived $\mu^{\mathrm{true}}$. $\chi^2(\alpha,\beta,M_\mathrm{B})$ in the Eq.
(\ref{chi2}) has a minimum at $M_{\mathrm{B}}=B/C$~\cite{Goliath2001}, and it is
\begin{equation}\label{chi2hat}
\widetilde{\chi}^2(\alpha, \beta)=A(\alpha, \beta)-\frac{B(\alpha, \beta)^2}{C(\alpha, \beta)}.
\end{equation}
Therefore, by minimizing $\widetilde{\chi}^2(\alpha, \beta)$, we can achieve calibrations for
$\alpha$ and $\beta$ with true distances derived from cosmic chronometer observations.
On the other hand, we also can benefit an estimation for the nuisance parameter, $M_\mathrm{B}$,
from this fitting. It should be noted that this merit enable us to break the degeneracy between
the Hubble constant $H_0$ and the absolute magnitude $M_\mathrm{B}$ when we investigate cosmological
implications of this Hubble diagram in the following analysis.

\section{Results}

The light-curve fitting parameters calibrated from the reconstructions of $H(z)$ with
different prior mean functions for GP and initial guess models for NPS taken into
consideration are summarized in Table~\ref{Tab1} and Table~\ref{Tab2}, respectively. As shown
in Table~\ref{Tab1}, it is suggested that GP reconstructions and following calibrations and
cosmological implications are somewhat sensitive to the assumption of the prior mean function. However,
from the Table~\ref{Tab2}, it is found that NPS reconstructions and following results are
not dependent on the initial guess model. The difference
between the distance modulus $\mu(z)$ derived from the reconstructed $H(z)$ functions with the
GP and NPS presented in Figure 1a, and the $\Lambda$CDM values is shown in Figure 2. In agreement
with the results of Figure 1b, the NPS values result in a Hubble diagram
more consistent with the $\Lambda$CDM prediction than do the GP
values. Here, again, it should be noted that, rather than fixing $H_0=70
~\mathrm{km}~\mathrm{s}^{-1}~\mathrm{Mpc}^{-1}$ to obtain the values
of $M_\mathrm{B}^1$ and $\Delta_M$ in the global fit for the $\Lambda$CDM
model~\cite{Betoule14}, these values, in our analysis, can be estimated from the
cosmological-model-independent calibration by confronting
light-curve fitting parameters-dependent distances of
SNe Ia (Eq.~\ref{eq1}) with the ones derived from the reconstructed function of
$H(z)$.

\begin{figure}[h]
\label{Fig2}
\includegraphics[width=0.49\textwidth, height=0.44\textwidth]{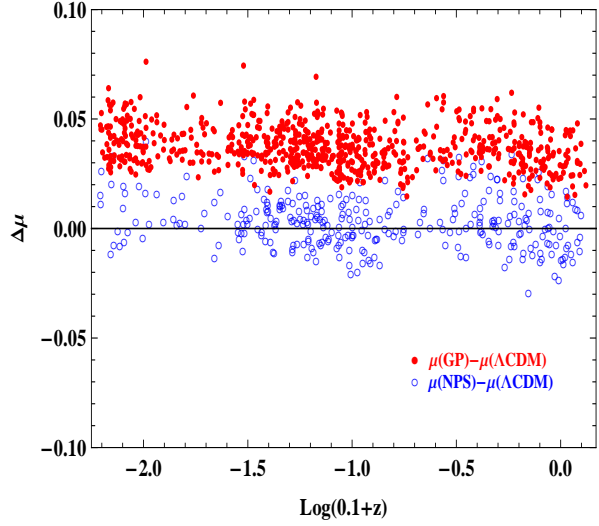}
\caption{The estimated distance modulus difference calculated with
the GP, NPS and the $\Lambda$CDM values obtained from the
light-curve fitting parameters.}
\end{figure}

We also investigate some cosmological implications of these
model-independent Hubble diagrams for the JLA sample.  Assuming a
spatially flat $\Lambda$CDM scenario, whose expansion rate is given
by $H(z)=H_0[\Omega_m(1+z)^3+(1-\Omega_m)]^{1/2}$, where $H_0$
and $\Omega_m$ are the present value of the Hubble parameter and matter
density parameter, respectively, we derive the constraints
on the $\Omega_m-H_0$ plane and the results are also summarized in Table~\ref{Tab1}
and Table~\ref{Tab2}. Once again results based on GP are slightly dependent on the
prior mean function. In contrast, results based on NPS are hardly sensitive
the initial guess model. Moreover, we also present constrained contours
on the $\Omega_m-H_0$ plane with reconstructed functions in the Figure 1a taken into
account. For the Gaussian processes, we obtain
that the value of the Hubble constant is fairly compatible with the
constraint from the latest $Planck$+WMAP9+BICEP2 CMB
measurements reported in Ref.~\cite{Bin14}. For the smoothing
method, we found that the constraint on the Hubble constant is excellently consistent
with the recent measurement of the local
Hubble parameter obtained from the recession velocity of objects
around us ($H_0=73.80 \pm 2.40$ at 68.3\% confidence
level)~\cite{Riess11} (see Figure 3b).

\begin{figure*}
\includegraphics[width=0.43\textwidth, height=0.447\textwidth]{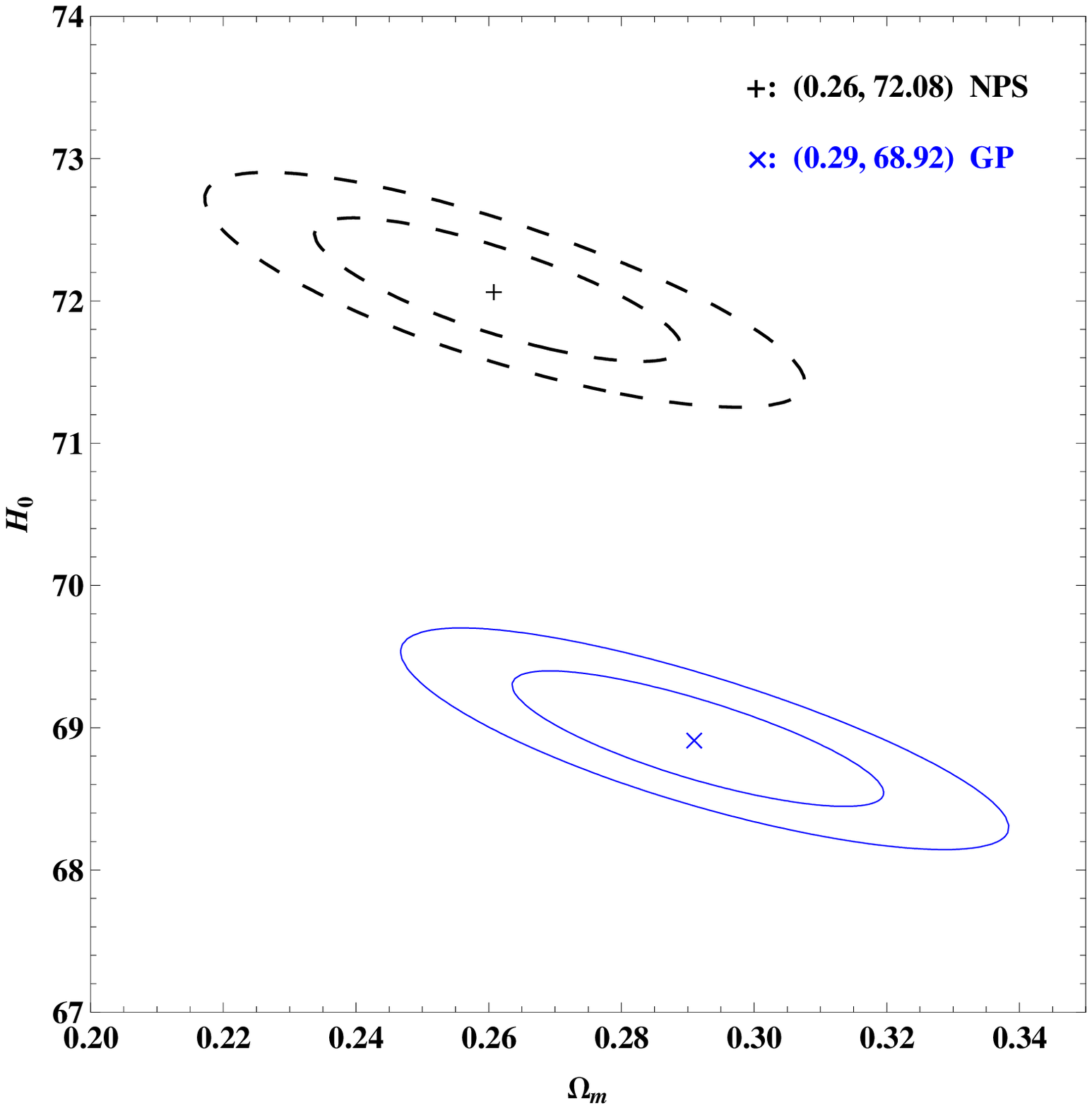}
\hspace{0.3cm}
\includegraphics[width=0.50\textwidth, height=0.45\textwidth]{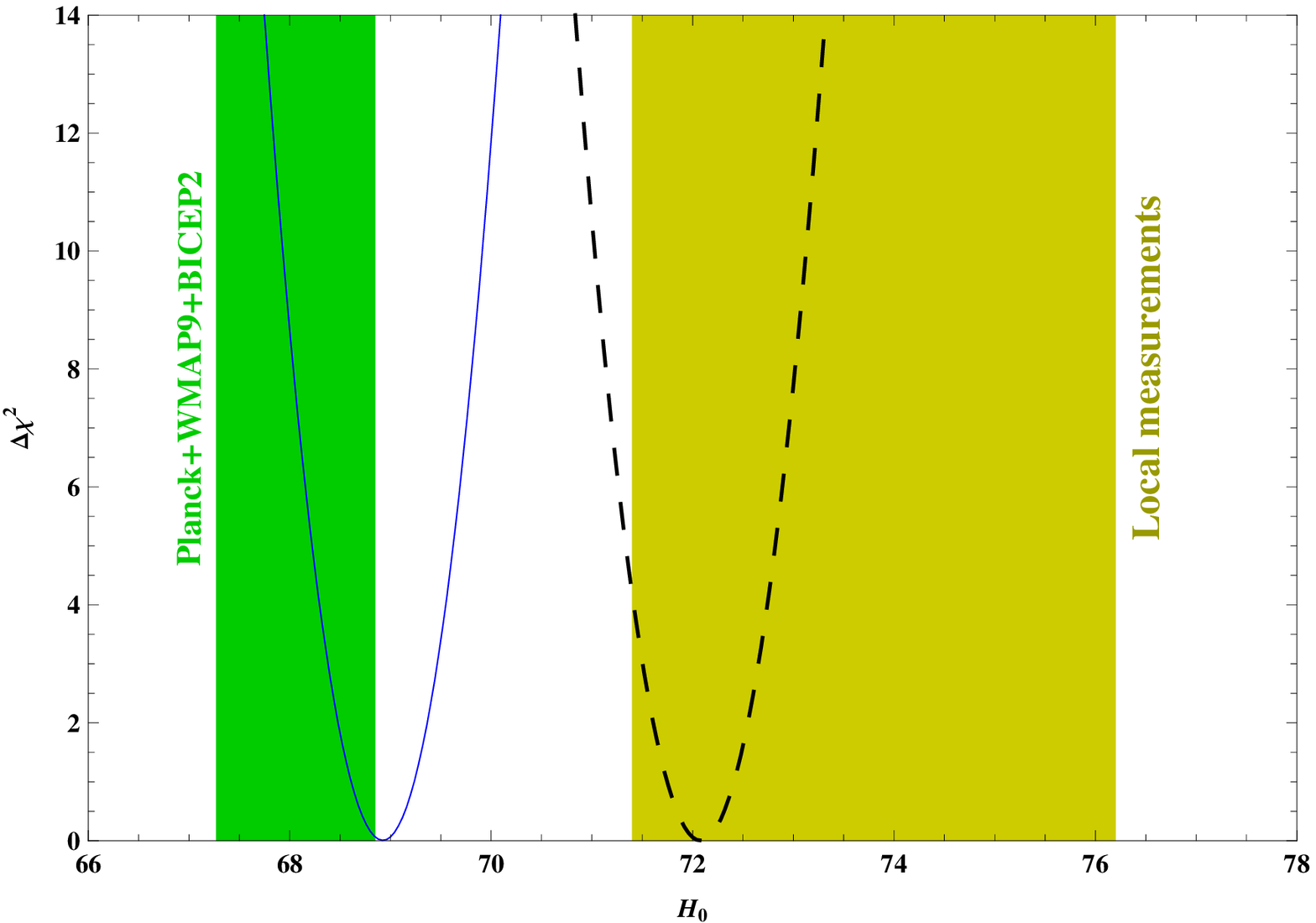}
\caption{ {\bf Left:} Constraints on the spatially flat $\Lambda$CDM
scenario using the model-independent Hubble diagrams of the JLA
compilation. {\bf Right:} Marginalized distribution of the local
Hubble parameter $H_0$ from the $H(z)$-calibrated Hubble diagrams of
JLA sample using Gaussian processes (blue solid) and the
non-parametric smoothing method (black dashed). The dark green and
dark yellow rectangles correspond to the 68.3\% interval of $H_0$
obtained  from the latest $Planck$+WMAP9+BICEP2 CMB~\cite{Bin14} and
Cepheids observations~\cite{Riess11}, respectively.}
\end{figure*}

\begin{table}[!h]
\begin{tabular}{|c|c|c|}
\hline
~Prior mean functions ~&~~Calibrations on ($\alpha, \beta, M, \Delta M$)~~&~~Constraints on $\Omega_m$ and
$H_0$~~\\
\hline
Constant~~&~~0.137,~3.036,~-19.081,~-0.056~~&~~$0.291^{+0.029}_{-0.027},~68.924^{+0.476}_{-0.477}$~~\\
\hline
E-D~~&~~0.149,~3.157,~-19.042,~-0.029~~&~~$0.270^{+0.028}_{-0.027},~70.838^{+0.493}_{-0.494}$~~\\
\hline
$\Lambda$CDM~~&~~0.158,~3.439,~-18.983,~-0.031~~&~~$0.255^{+0.029}_{-0.028},~72.795^{+0.521}_{-0.522}$~~\\
\hline
wCDM~~&~~0.152,~3.307,~-19.053,~-0.042~~&~~$0.267^{+0.029}_{-0.028},~70.216^{+0.497}_{-0.498}$~~\\
\hline
\end{tabular}
\tabcolsep 0pt \caption{\label{Tab1} Summary of the results with different prior mean functions for the GP.} \vspace*{5pt}
\end{table}

\begin{table}[!h]
\begin{tabular}{|c|c|c|}
\hline
~Initial guess models ~&~~Calibrations on ($\alpha, \beta, M, \Delta M$)~~&~~Constraints on $\Omega_m$ and
$H_0$~~\\
\hline
E-D~~&~~0.154,~3.190,~-19.022,~-0.019~~&~~$0.264^{+0.028}_{-0.027},~71.738^{+0.500}_{-0.501}$~~\\
\hline
$\Lambda$CDM~~&~~0.155,~3.221,~-19.014,~-0.017~~&~~$0.261^{+0.028}_{-0.027},~72.070^{+0.504}_{-0.505}$~~\\
\hline
wCDM~~&~~0.153,~3.172,~-19.028,~-0.021~~&~~$0.265^{+0.028}_{-0.027},~71.525^{+0.498}_{-0.499}$~~\\
\hline
\end{tabular}
\tabcolsep 0pt \caption{\label{Tab2} Summary of the results with different initial guess models for the NPS.} \vspace*{5pt}
\end{table}

\section{Conclusions}

As is well known, implications derived from current SNe Ia analyses,
where the light-curve fitting parameters are usually determined to the
global fit in the frame of the standard dark energy model, are
cosmological-model-dependent. In this paper, we have applied two non-parametric
methods to reconstruct the Hubble
expansion using 15 $H(z)$ measurements ($z \leq 1.2$) from cosmic
chronometers and transformed the reconstructed functions into
distances by carrying out a numerical integration. The choice of
this reduced $H(z)$ sample is based on the arguments of
Refs.~\cite{morescoetal, licia2014}, which ensures that the
evolution of the Hubble parameter reconstructed in our analysis is
neither dependent on the cosmology nor on the stellar population
model.

By using the derived model-independent distances we have calibrated
the light-curve fitting parameters and constructed a completely
cosmological model-independent Hubble diagram for the JLA sample.
The results suggest that the uncertainties on the light-curve
fitting parameters obtained from the $H(z)$ reconstructions are
almost of the same order of magnitude as the ones determined in the
global fit for the $\Lambda$CDM model. Therefore, we expect the
constraining power of any analysis derived from these
$H(z)$-calibrated Hubble diagrams to be nearly identical to the one
obtained when the global fit to a given model is performed. It
should be emphasized, however, that cosmological implications of the
Hubble diagrams constructed from $H(z)$ data do not suffer with
cosmological model-dependence.

However, we  have shown that these diagrams and their implications
depend reasonably on the method used to reconstruct the Hubble
evolution. Furthermore, for Gaussian processes, reconstruction and following
analysis are obviously sensitive to the prior mean function. Assuming the
spatially flat $\Lambda$CDM model, we have derived constraints on the matter
density parameter $\Omega_m$ and Hubble constant $H_0$ from the JLA sample.
For the analysis that uses Gaussian processes, it is shown that constraints on
model parameters, $\Omega_m$ and $H_0$, vary obviously when different
prior mean functions are considered. In the reasonable and safe case with
the prior mean function being constant, it is shown that the constraint on
$H_0$ is quite compatible with that derived from the
latest Planck+WMAP9+BICEP2 CMB observations at 68.3\% confidence level.
When the non-parametric smoothing procedure is applied, results are hardly dependent
on the initial guess model and consistently favor a higher value of $H_0$, which are
in excellent agreement with the recent local measurement of the expansion rate from
Cepheids observations.

A final aspect worth emphasizing is that, differently from the GP
reconstruction, the expansion rate $H(z)$ obtained from the
smoothing method behaves similarly to the one predicted by the
$\Lambda$CDM model, with the deceleration/acceleration transition
happening around $z \simeq 0.6$ (Figure 1b). Therefore, taking the
standard evolution as a good description for the late-time evolution
of the Universe, the NPS results obtained in this analysis seem to
be more consistent than those derived from the GP method.

\section*{Acknowledgments}
Zhengxiang Li, Hongwei Yu, and Zong-Hong Zhu are supported by the
Ministry of Science and Technology National Basic Science Program
(Project 973) under Grants Nos. 2012CB821804 and 2014CB845806, the
National Natural Science Foundation of China under Grants Nos.
11505008, 11373014, 11073005, 11375092, and 11435006, the China Postdoc Grant
No. 2014T70043, and the Youth Scholars Program of Beijing Normal
University. J. E. Gonzalez and J. S. Alcaniz are supported by CAPES,
CNPq and FAPERJ (Brazilian Agencies).

\end{document}